\documentclass{aa} 

\usepackage{graphicx}
\usepackage{txfonts}
\usepackage{textcomp}

\usepackage{natbib}
\bibpunct{(}{)}{;}{a}{}{,} 

\catcode`\@=11
\def\gsim{\ifmmode{\mathrel{\mathpalette\@versim>}}
    \else{$\mathrel{\mathpalette\@versim>}$}\fi}
\def\lsim{\ifmmode{\mathrel{\mathpalette\@versim<}}
    \else{$\mathrel{\mathpalette\@versim<}$}\fi}
\def\@versim#1#2{\lower 2.9truept \vbox{\baselineskip 0pt \lineskip 
    0.5truept \ialign{$\m@th#1\hfil##\hfil$\crcr#2\crcr\sim\crcr}}}
\catcode`\@=12

\begin{document}

   \title{The Giraffe Inner Bulge Survey (GIBS) II. Metallicity distributions and alpha element abundances at fixed Galactic latitude
   \thanks{Based on observations taken with ESO telescopes at the La Silla Paranal 
   Observatory under programme IDs 187.B-909(A) and 71.B-0196.}}

   \subtitle{}

\author{
 O. A. Gonzalez\inst{1}
\and
 M. Zoccali\inst{2,3} 
 \and
 S. Vasquez\inst{2,3}
 \and 
 V. Hill\inst{4}
 \and
 M. Rejkuba\inst{5, 6}
 \and           
 E. Valenti\inst{5}
 \and
A. Rojas-Arriagada\inst{4}
\and 
A. Renzini\inst{7}
 \and
 C. Babusiaux\inst{8}   
 \and 
 D. Minniti\inst{3,9,11}
\and
 T. M. Brown\inst{10}         
}

\institute{
European Southern Observatory, A. de Cordova 3107, Casilla 19001, Santiago 19, Chile.
\email{ogonzale@eso.org}
\and
Instituto de Astrof\'{i}sica, Facultad de F\'{i}sica, Pontificia Universidad Cat\'olica 
de Chile, Av. Vicu\~na Mackenna 4860, Santiago, Chile.
\and
Millennium Institute of Astrophysics, Av. Vicu\~{n}a Mackenna 4860, 782-0436 Macul, 
Santiago, Chile.
\and
Laboratoire Lagrange UMR 7293, Universit\'e de Nice Sophia-Antipolis, CNRS, 
Observatoire de la C\^{o}te d'Azur, CS 34229, 06304 Nice, cedex 04, France.
\and 
European Southern Observatory, Karl-Schwarzschild-Strasse 2, 85748, Garching, Germany.
\and
Excellence Cluster Universe, Boltzmannstr. 2, 85748, Garching, Germany
\and
INAF - Osservatorio Astronomico di Padova, vicolo dell'Osservatorio 5, 35122, Padova, 
Italy.
\and
GEPI, Observatoire de Paris, CNRS UMR 8111, Universit\'e Paris Diderot, F-92125, Meudon,
Cedex, France.
\and
Departamento de Ciencias F\'isicas, Universidad Andr\'es Bello, Rep\'ublica 220, Santiago, Chile
\and 
Space Telescope Science Institute, 3700 San Martin Drive, Baltimore, MD 21218, USA.
\and
Vatican Observatory, V00120 Vatican City State, Italy.
}
   \date{Received 15 Jun, 2015; accepted Aug 03, 2015}

 
  \abstract
   {} 
   {We investigate metallicity and $\alpha$-element abundance gradients 
along a Galactic longitude strip, at latitude $\rm b\sim-4^{\circ}$, with the aim of providing 
observational constraints for the structure and origin of the Milky Way 
bulge.}
   { High resolution (R$\sim$22,500) spectra for 400 K giants, in four fields
     within $\rm -4.8^{\circ} \lesssim b \lesssim -3.4^{\circ}$ and  $\rm -10^{\circ} \lesssim l \lesssim +10^{\circ}$, were obtained within
     the GIRAFFE Inner Bulge Survey (GIBS) project. To this sample we added
     another $\sim$ 400 stars in Baade's Window at $\rm (l,b)=(1^{\circ},-4^{\circ})$,
     observed with the identical instrumental configuration: FLAMES GIRAFFE in
     Medusa mode with HR13 setup. All target stars lie within the red clump
     of the bulge color magnitude diagram, thus minimizing contamination from
     the disc or halo stars. The spectroscopic stellar surface parameters were
     derived with an automatic method based on the 
GALA code, while the [Ca/Fe] and [Mg/Fe] abundances as a function of 
[Fe/H] were derived through a comparison with the synthetic spectra using MOOG.
We constructed the metallicity distributions for the entire sample, as well 
as for each field individually, in order to investigate the presence of gradients 
or field-to-field variations in the shape of the distributions.
}
   { The metallicity distributions in the five fields are consistent with being
     drawn from a single parent population, indicating the absence of a gradient
     along the major axis of the Galactic bar. The global metallicity distribution
     is well fitted by two Gaussians. The metal poor component is rather broad,
     with a mean at $\rm <[Fe/H]>=-0.31$ dex and $\sigma=0.31$ dex. The metal-rich one
     is narrower, with mean $\rm <[Fe/H]>=+0.26$ and $\sigma=0.2$ dex.
     The [Mg/Fe] ratio follows a tight trend with [Fe/H], with enhancement with
     respect to solar in the metal-poor regime, similar to the one observed
     for giant stars in the local thick disc. [Ca/Fe] abundances follow a similar trend, but with a considerably larger scatter than [Mg/Fe]. A decrease in [Mg/Fe] is observed at $\rm [Fe/H]=-0.44$ dex. This \textit{knee} is in agreement with our previous bulge 
study of K-giants along the minor axis, but is 0.1 dex lower in 
metallicity than the one reported for the Bulge microlensed dwarf and sub-giant stars. We found 
no variation in $\alpha$-element abundance distributions between different 
fields.}
   {}

\titlerunning{The Giraffe Inner Bulge Survey (GIBS) II}
   \maketitle
%
\section{Introduction}

In the last few years we have witnessed important progress towards the understanding of the Milky Way bulge. The availability of large photometric and spectroscopic datasets from ongoing surveys are finally allowing us to obtain a wider view of the Bulge properties, even expanding our knowledge to the innermost regions previously not accessible. 

The bar in the inner regions of the Galaxy was first suggested by \citet{deVac64} and its main structural properties have been investigated in detail since then \citep[][ and  references therein]{blitz+91, stanek+94, dwek+95,   babusiaux+05,   rattenbury+07, gonzalez+11}. The use of red clump (RC) giant stars as a distance indicator has been a fundamental tool for Bulge morphology studies \citep{stanek+94}. Based on this technique the axial ratios of the bar have been currently constrained to be about 1:0.4:0.3 with a bar size of about 3.1$-$3.5 kpc major-axis length. The position angle of the bar has been historically measured to range between $\sim$20--40 deg. with respect to the Sun-centre line of sight with its near end towards positive Galactic longitudes \citep{blitz+91, stanek+94, dwek+95, binney+97, bissantz-gerhard+02, benjamin+05, babusiaux+05, rattenbury+07, cao+13}. Recently, detailed comparison of RC stellar counts to N-body models \citep[e.g.][]{wegg-gerhard+13} have allowed to constrain the bar position angle to $27-33^{\circ}$.
The RC technique to measure distances towards the Bulge led to the discovery of a split RC at $\rm l=0^\circ$  and  $\rm |b|>5^\circ$ by \citet{nataf+10} and \citet{mcwilliam-zoccali+10}. This discovery was quickly followed by the construction of wider, more detailed 3D maps that revealed the X-shaped structure of the Bulge \citep{saito+11,  wegg-gerhard+13} as first suggested by \citet{mcwilliam-zoccali+10}. This X-shape structure is often seen in external galaxies and is well reproduced by dynamical models of disc galaxies \citep[e.g.][]{li-shen+12,gardner+14}. They correspond to an extreme case of the boxy/peanut (B/P) structures observed in several external disc galaxies formed as the consequence of buckling instabilities of galaxy bars. The buckling process of the bar results in the heating up of stellar orbits in the vertical direction \citep[e.g.,][]{ComDebFri90, Ath05,  MarShlHel06, debattista+05}. Due to the absence of an external influence in this process, a reference of these structures as so-called pseudo-bulges \citep[see][for a detail definition of such structures]{KorKen04} is often found in the literature. However, we refrain from using such a definition to differentiate them from the young stellar disc-like structures formed in the inner part of bars due to the continuous in-fall of gas. We will then refer to this structure simply as the B/P bulge of the Milky Way.

The Bulge Radial Velocity Assay survey \citep[BRAVA;][]{howard+09, kunder+12}, measured radial velocities for $\sim$10,000 Bulge M giants at latitudes $\rm b=-4^\circ,-6^\circ,-8^\circ$ and longitudes  $\rm -10^\circ<l<10^\circ$. The BRAVA survey provided strong evidence for the cylindrical rotation of the Bulge, concluding that a B/P bulge would be enough to reproduce the rotation curve of the Bulge without the need of a  {\it classical} component, formed via mergers in the early evolution of the Galaxy \citep{howard+09, shen+10}. The cylindrical rotation was also confirmed recently by the GIBS survey in \citet[][hereafter Paper I]{zoccali+14} where it was extended to latitudes ($\rm b\sim-2^\circ$). 

While the B/P bulge has been shown to rotate cylindrically, the bulk of its stellar population is over $\sim$10 Gyr old \citep[e.g.][]{zoccali+03,clarkson+11,valenti+13}. A $\sim$10 Gyr look-back time brings us to redshift z$\sim$2, where galaxies are radically different from local ones and live at the epoch when the overall star formation rate (SFR) in the Universe peaked. Massive galaxies with stellar masses comparable to those seen today ($\rm \sim10^{10}-10^{11}$Msun) are forming stars at rates of $\sim$20-200 Msun/yr, some 20 times higher than in local galaxies of the same mass. These
higher SFRs are a consequence of higher gas mass fractions in these systems, typically $\sim$30-50\% as revealed by
CO observations \citep{tacconi+10, tacconi+13, daddi+10}. Moreover, adaptive optics resolved H$\alpha$
kinematic studies have revealed that most z$\sim$2 main-sequence galaxies are large rotation-dominated discs, with a minor fraction of major mergers, are characterised by very high velocity dispersion ($\sim$50-100 km/s) and by the presence of several kpc scale, actively
star-forming clumps \citep[e.g.][]{forster-schreiber+09, genzel+11, mancini+11, newman+13}. In parallel with these findings, a new paradigm has emerged in which bulges form by the migration and coalescence to the center of massive star-forming clumps \citep{ImmSamGer04, BouElmElm07, elmegreen+08,bournaud+09}, a situation which may even develop into a global violent disc instability, leading to a dissipational
formation of compact bulges \citep{dekel-burkert+14}. Considering all the process that can be involved in the formation of bulges, it becomes clear that kinematics alone would not be sufficient to reconstruct the formation mechanism of the Galactic bulge.  Indeed, the chemical abundance measurements of individual stars have been shown to provide a different perspective, especially when combined with the kinematics.

Several studies have investigated the metallicity distribution functions near to Baade's Window at $\rm (l,b)=(0^\circ,-4^\circ)$ \citep[e.g.][]{rich+88, mcwilliam-rich+94, minniti+96, sadler+96, ramirez+00, zoccali+03, fulbright-mcwilliam-rich+06} deriving a wide distribution with metallicities ranging from -1.6 to 0.5 dex and a peak at solar metallicities. \citet{zoccali+08}, and \citet{johnson+11, johnson+13} used high resolution spectra to derive metallicity distributions in different Bulge regions along the minor axis and have firmly established the presence of a vertical metallicity gradient with field stars closer to the Galactic center being on average more metal-rich. Furthermore, the general picture of these metallicity gradients in the Bulge was presented in \citet{gonzalez+13} who showed the global mean photometric metallicity map of the Galactic bulge based on the Vista Variables in the Via Lactea (VVV) ESO public survey data \citep{minniti+10}.
 
By combining  [Fe/H]  and   [Mg/Fe]   abundances  and   kinematics, \citet{hill+11} and \citet{babusiaux+10}  suggested  the  presence  of two distinct  components  in  the  Bulge:  a   metal  poor  one ($\rm <[Fe/H]>\sim-0.3$)  with  kinematics typical of a classical spheroid, and a metal rich one,  ($\rm <[Fe/H]>\sim+0.3$)  concentrated  towards the  Galactic plane with a significant vertex deviation, suggestive of a bar-like component \citep[see also][]{soto+07}. The origin of the observed metallicity gradients would then be the natural consequence of the different contribution of each of these components as a function of Galactic latitude. A similar conclusion was reached by \citet{rojas-arriagada+14} based on the Gaia-ESO survey data.

\begin{table*}[]
\caption{Observed fields and final number of stars analysed in this work.}
\label{log}
\centering
{\small
\begin{tabular}{l c c r r c c c r r r r}
\hline\hline
Field name & RA & DEC & l & b & Setup & R=$\lambda/\Delta \lambda$ & $\lambda$ coverage & N$_{\rm stars}$ & exptime/star & $\rm A_V$\\  
       & (hr) & (deg) & (deg) & (deg) & &                          &   $\AA$            & GIRAFFE       &  (s) & mag \\       
\hline
\\
HRp8m3 & 18:16:40.8 & -23:45:32.20 &   7.9460 & -3.4770    & HR13 & 22500 & 6120-6405 & 103 & 27000 & 1.71   \\
HRp4m3 & 18:07:15.4 & -27:31:21.70 &   3.6174 & -3.4111    &  "       &  "         &     "              & 88   & 27000 & 1.61   \\
HRm5m3 & 17:47:49.2 & -35:03:24.10 & 355.0036 & -3.5701 &  "       &  "         &     "              & 103 & 27000 & 1.59   \\
HRm7m4 & 17:48:11.0 & -37:09:25.30 & 353.2336 & -4.7106 &  "       &  "         &     "              & 106 & 27000 & 1.33   \\
\\
\hline
\end{tabular}
}
\end{table*}

A step further in the interpretation of the Bulge metallicity distribution was done using the Abundances and Radial velocity Galactic Origins Survey \citep[ARGOS;][]{freeman-argos+13}. \citet{ness-abu+13,  ness-kine+13} measured radial velocities, [Fe/H], and [$\alpha$/Fe] ratios for $\sim$28,000 Bulge stars at different regions well-distributed across the Bulge. They find the same cylindrical rotation found by the BRAVA and GIBS survey, but suggested the presence of 5 components in the metallicity distribution of the Bulge. They suggested that the actual Bulge component would be the metal rich one ($\rm [Fe/H]\sim+0.15$), and perhaps the component at $\rm [Fe/H]\sim-0.25$ dominating at high latitudes. Additional components would belong to the inner disc and halo. 

On the other hand, $\alpha$-element abundances in Bulge stars provide us with with an additional constraint on the formation history of the Bulge stellar populations: its formation time-scale. As suggested by \citet{tinsley+79} the ratio of [$\alpha$/Fe] compared to [Fe/H] is expected to be a function of the time delay between the production of both $\alpha$- and iron-peak elements by SNe II \citep[e.g.][]{woosley+95} and the yield of mostly iron-peak elements with little $\alpha$-element production by SNe Ia \citep[e.g.][]{nomoto+84}. Clearly, a sufficient amount of time needs to be accounted for such that enough SNe Ia events occur for the [$\alpha$/Fe] ratio to decline from the SNe II value.  The $\alpha$-element abundances of Bulge stars with [Fe/H]$<$-0.3 have been historically found to be enhanced over iron by [$\alpha$/Fe]$\sim$+0.3 dex \citep{McWRic94, rich-origlia+05, cunha-smith+06, fulbright-mcwilliam-rich+07, lecureur+07, rich-origlia-valenti+07} thus suggesting a fast formation scenario.  Metal-rich stars on the other hand show a decrease in [$\alpha$/Fe] reaching [$\alpha$/Fe]$=0$ for metallicities above Solar values. However, the SNe Ia delay time might vary depending on the different production channels that could be present, and therefore to obtain a direct translation of these trends to absolute time scales is not straightforward. As a consequence, the comparison of [$\alpha$/Fe] values in Bulge stars against those of other galactic components has been proven to be a useful tool to investigate the Bulge formation time-scale with respect to those other components such as disc and halo. Based on this relative measurement approach, \citet{fulbright-mcwilliam-rich+07}, \citet{zoccali+06}, and \citet{lecureur+07} concluded that the [$\alpha$/Fe] ratio was enhanced by nearly +0.1 dex with respect to the trends of both the local thin and the thick disc as traced by nearby dwarf stars. Therefore, these results implied a shorter formation time scale for the Bulge than from both discs. However, \citet{melendez+08}, \citet{alves-brito+10}, \citet{johnson+11}, \citet{johnson+13}, and \citet{gonzalez+11} found a similarity between the [$\alpha$/Fe] abundance ratio of Bulge and thick disc giant stars. The origin of this discrepancy has been pointed to a systematic difference when analysing the abundances of dwarf stars from the disc and Bulge giants \citep{melendez+08}. In support of this scenario, the microlensed dwarfs from the Bulge have shown the same similarity in the $\alpha$-element enhancement to those of the thick disc dwarfs \citep{bensby+11}. Therefore, the usage of well-calibrated measurements on homogenous datasets are fundamental for these kinds of studies.

A few high-resolution studies have looked at a few fields located at relatively similar latitude between each other \citep[e.g.][]{johnson+13}, but always at latitudes higher than $\rm b=-4^\circ$, where the contribution of the metal-rich stars is known to decrease \citep{zoccali+08,johnson+13,ness-abu+13}. The limited samples could therefore miss some subtle gradients. This is now investigated in this work with $\sim$100 stars per field in a region where the two 
populations are well represented. This study provides important constraints on the global view of the metallicity and $\alpha$-element distribution of the Bulge which are fundamental to reconstruct its formation scenario. 

\section{GIBS high-resolution spectra}

The Giraffe Inner Bulge Survey (GIBS) is a survey of $\sim$ 6,500 RC stars in the Milky Way bulge observed with the GIRAFFE spectrograph of the FLAMES instrument \citep{pasquini+00} at the ESO Very Large Telescope  (VLT). The aim of the GIBS survey is to investigate the metallicity and radial  velocity distribution of Bulge stars across different fields, spread over a large  area of the inner Galactic bulge ($-10^{\circ}\leq l \leq +10^{\circ}$ and $-10^{\circ}\leq b \leq +4^{\circ}$). A detailed description of the target selection and data products of the program is given in Paper I. In this article, we focus on the analysis of the high-resolution spectra of 400 RC stars located in four fields at Galactic latitude $b\sim -3.5^{\circ}$ for which we provide a brief description summary.

\begin{figure}[]
\centering
\includegraphics[width=9cm, angle=0]{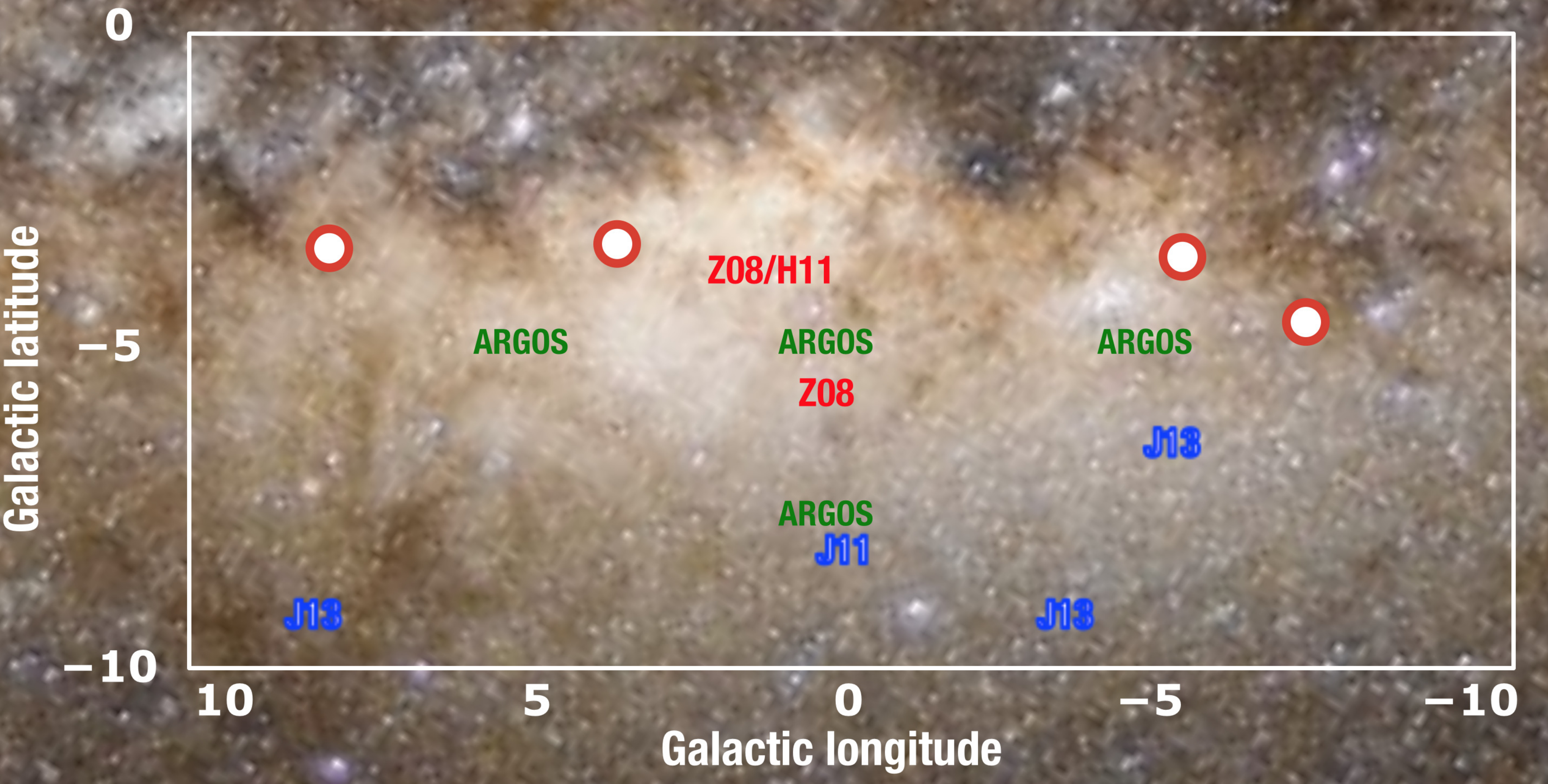}
\includegraphics[width=9cm, angle=0]{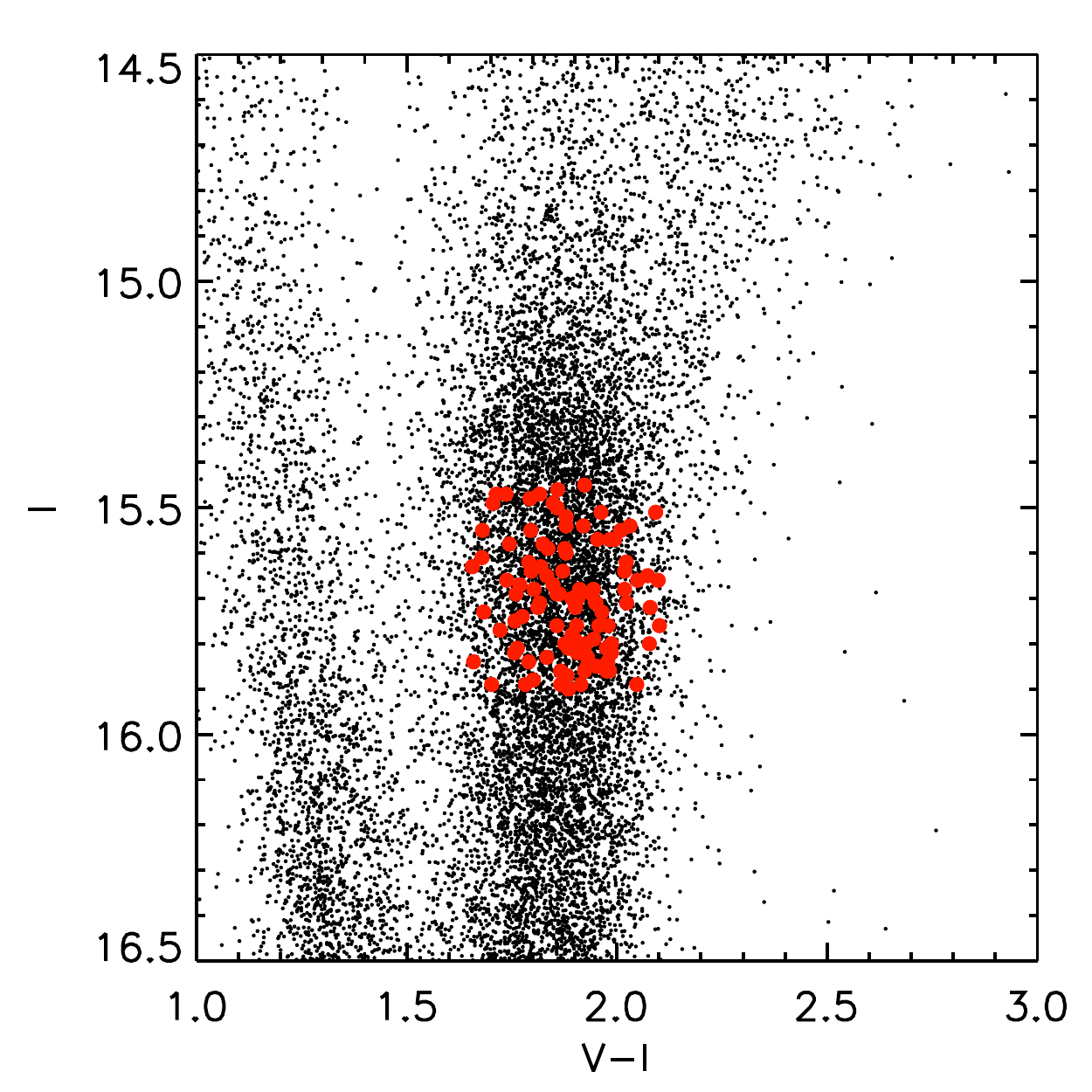}
\caption{The  location of  the fields analysed in the present study (red circles) overplotted on an optical image of the Milky Way bulge (\copyright Serge Brunier). The location of the fields discussed in previous studies \citep{zoccali+08, hill+11} are shown as Z08 and H11, as well as the location of \citet{ness-abu+13}, \citet{johnson+11}, and \citet{johnson+13} fields as ARGOS, J11 and J13, respectively. Lower panel shows the colour-magnitude diagram for field HRm5m3. Red filled circles show the corresponding target stars for this field.}
\label{VVVmap}
\end{figure}

Target stars for the spectroscopic observations were selected from the VVV multi-band photometric catalogues  \citep{gonzalez+12}. The location of the fields was carefully selected in order to overlap with additional optical photometric observations from the OGLEII survey \citep{sumi+04pm}. This allows us to add extra information on the target stars such as proper motions and a larger colour baseline ($\rm V-K_s$) for deriving the initial values for effective temperatures.  

Figure~\ref{VVVmap} shows the location of the 4 fields analysed in the present study overplotted on an optical image of the Milky Way bulge. The location of other fields with previous studies similar to the one presented here \citep{zoccali+08, ness-abu+13, johnson+11, johnson+13} are also shown. The colour-magnitude diagram for one of our observed fields (HRm5m3), where our selected targets are clearly highlighted, is also shown in Fig.~\ref{VVVmap}. The selection box of targets in the colour-magnitude diagram can be clearly seen in Fig.\ref{VVVmap} and it was designed to target RC stars while minimising the contamination for foreground disc stars. 

A description of the  observations is  reported in  Table~\ref{log}.  All spectra in the fields analysed in this study  were  observed using the high-resolution grating (R=22,500) through setup HR13 centered at $\sim 6300 \AA$ in order  to  measure  the  chemical  abundance of  iron  and  $\alpha$-elements in a similar way as in \citet{gonzalez+11}. 

Reduction of the spectra was carried out using the ESO GIRAFFE pipeline. Bias and flat-field correction, individual spectral extraction, and wavelength calibration have been applied by using the ESO/GIRAFFE pipeline\footnote{http://www.eso.org/sci/software/pipelines/}. An adequate sky spectrum for each exposure was obtained by co-addition of sky spectra obtained from dedicated fibres in each GIRAFFE configuration. Sky subtraction was carried out on each spectrum using \textit{skytweak} task in IRAF as described in Paper I. 
Heliocentric radial velocities were measured using IRAF \textit{fxcor} task by cross--correlation against a synthetic spectra for a typical bulge 
K giant star ($\rm T_{eff}=4500$K, log$g=2.3$ and $\rm [Fe/H]=-0.3$),  covering the corresponding wavelength range of the GIRAFFE HR13 setup from $6100 \AA$ to $6400 \AA$. For each star, multiple exposures ($\sim$10) were individually corrected for radial velocity and individual products were mean combined to produce the final set of spectra for each field. The errors measured on radial velocities were typically $\sim 1$ km/s.

\section{Stellar parameters}

Stellar parameters, namely effective temperature ($\rm T_{eff}$), surface gravity ($\rm \log{g}$), microturbulence velocity ($\rm \xi$) and metallicity ([Fe/H]) were derived based on an iterative method similar to the one described in \citet{zoccali+08} and \citet{gonzalez+11}. The method consists of the spectroscopic refinement of a set of the first guess photometric stellar surface parameters. The objective is to find the best combination of parameters in order to impose an excitation equilibrium of Fe lines (zero slope of Fe abundance as a function of excitation potential of Fe lines) and simultaneously measuring the same Fe abundance for all lines independently of their equivalent width (measuring a zero slope of Fe abundance as a function of $\rm \log(EW/\lambda)$). Note that we rely on the usage of only FeI lines for this procedure, as the resolution GIRAFFE spectra does not provide us with sufficient clean FeII lines across our analysed spectral region.  In \citet{zoccali+08} and \citet{gonzalez+11} this process was done manually, in the sense that the parameters were modified \textit{by hand} to minimise the corresponding slopes. This manual procedure was found to produce some systematics, particularly in the high metallicity regime \citep{hill+11}. To correct for this issue and bring our datasets into a common baseline, we have now improved our method into an automatic procedure, similar to the one used in \citep{hill+11} but using the GALA code \citep{GALA+13} to spectroscopically refine the first guess photometric parameters. Note that another difference with respect to the aforementioned studies is that for this work we have used ATLAS9 model atmospheres (Castelli \& Kurucz 2004). This is because GALA already includes a dynamic call to the ATLAS9 code so that whenever GALA needs to investigate a new set of atmospheric parameters, ATLAS9 is called and a new model atmosphere is computed.

As a first step, the photometric temperature is calculated for each star using the  $\rm (V-K_s)$ colours from OGLEII  \citep{udalski+02} and VVV survey catalogues \citep{gonzalez+12}, dereddened based on the high resolution extinction maps from \citet{gonzalez+12}, and applying the \citet{ramirez+05} calibration.  Absolute V band magnitudes, calculated from the measured distances to each specific field from \citet{gonzalez+13} and bolometric corrections from \cite{alonso+99}, are then used to estimate photometric gravities based on the usual formula:

\begin{footnotesize}
\[ \rm
\log\left(g\right)=\log\left(g_{\odot}\right)+\log\left(\frac{M_*}{M_{\odot}}
\right)+0.4\left(M_{Bol,*}-M_{Bol,\odot}\right)+4\log\left(\frac{T_{eff,*}}{T_{eff,\odot}}\right)
\]
\end{footnotesize}

where  $\rm   M_{Bol,\odot}=4.72$,  $\rm   T_{eff,\odot}=5770$\,K, and
$\log\left(g_{\odot}\right)=4.44$ dex. A fixed value of $\rm M_*=0.8 M_{\odot}$ has been adopted similarly to \citet{zoccali+08} and \citet{hill+11}. 
Microturbulence velocity and global metallicity are set to 1.5 and 0.0, respectively, as a first step. These values are used to obtain a first guess ATLAS9 stellar model atmosphere and are subsequently refined spectroscopically by GALA using the equivalent widths of isolated Fe lines obtained by means of DAOSPEC \citep{DAOSPEC+08}.

After feeding GALA with the photometric stellar surface parameters and the corresponding model atmosphere, the code iteratively searches for spectroscopic effective temperatures and microturbulence velocity by imposing excitation equilibrium and the null slope of iron abundance versus equivalent width of the Fe lines.  During each iteration, a new model atmosphere is generated by GALA using the refined stellar parameters. Although available in GALA, we do not refine $\rm \log{g}$ values by requiring ionisation equilibrium, but we rely on the values derived photometrically. As discussed in \citet{zoccali+08}, the resolution of GIRAFFE is not high enough to resolve a sufficient number of clean FeII lines in our spectra and thus using the few available lines would introduce more errors than just using the photometrically derived values for $\rm \log{g}$. Therefore, after a first set of best-fitting values for $\rm T_{eff}$, $\rm \xi$, and [Fe/H] is found by GALA, we re-calculate the  $\rm \log{g}$ value according to the new parameters and perform another set of GALA iterations to refine the stellar parameters with the new  $\rm \log{g}$. 

\begin{figure}[]
\centering
\includegraphics[width=9.7cm, angle=0,trim=1.6cm 0.8cm 0cm 0.5cm,clip=true]{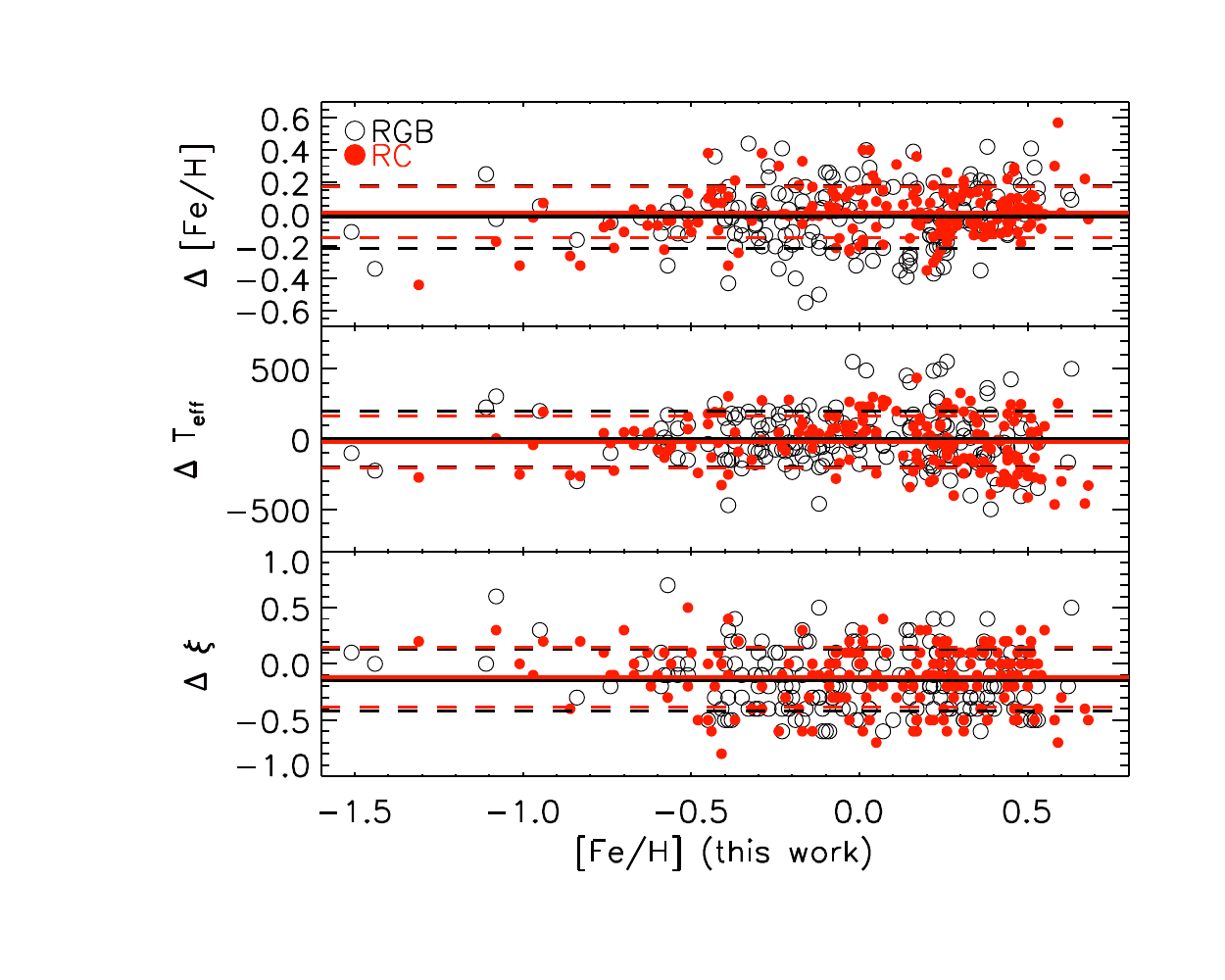}
\caption{Difference between metallicity (upper panel), effective temperature (middle panel) and microturbulence velocity (lower panel) for a sample of RC (red filled circles) and RGB (black empty circles) derived in this work and as derived in \citet{hill+11}. Solid lines in each panel indicate the mean difference of each sample and the dashed lines shows the 1$\rm \sigma$ range around the mean.}
\label{parameters}
\end{figure}

The list of FeI lines used for this method is the one used in Lecureur et al. (2007) following a careful cleaning of blended lines. The check for blends of the Fe lines was done using a spectrum of $\mu$ Leo obtained at the Canada-France-Hawaii Telescope with the ESPaDOnS spectro-polarimeter at a resolution R $\sim$80,000 and high S/N per pixel ($\sim$500). As a first step, the resolution of the $\mu$ Leo spectrum was degraded to the resolution of GIRAFFE HR13 setup of R$\sim$22,500. This was followed by a line-by-line inspection performed over the wavelength range of our science data. Several blended Fe lines were removed from the list ending with a total of 27 non-blended Fe lines in the HR13 setup. The $\log gf$ values from the original line list from \citet{lecureur+07} are calibrated so that each line provides an abundance of 0.30 dex from the EW measured in the spectrum of $\mu$ Leo while adopting $\rm T_{eff}=4540$ K, $\rm \log{g}=2.30$, and $\rm \xi=1.30$. Therefore, in order to further check our final Fe line list, we measured the EW of the corresponding line using DAOSPEC in the $\mu$ Leo spectrum (now at GIRAFFE resolution) and used GALA to derive the Fe abundance by fixing the model to those same stellar parameters from \citet{lecureur+07}. GALA retrieved an abundance of [Fe/H]=0.32, in excellent agreement with the literature value of [Fe/H]=0.34 \citep{gratton-sneden+90}. On the other hand, by leaving the stellar parameters free with the exception of $\rm \log{g}$ which was fixed to $\rm \log{g}=2.30$, GALA retrieves $\rm T_{eff}=4490$ K, $\rm \xi=1.43$ and [Fe/H]=0.31 for $\mu$ Leo. These values are in good agreement with those reported in the literature \citep{smith+00,gratton-sneden+90}. The final list of FeI lines used in our analysis is provided in Table~\ref{lines}.

Particular care has been devoted to ensure that the adopted stellar parameters were i) sufficiently accurate to allow reliable abundance estimates, and ii) consistent with previous analysis to guarantee a homogeneous comparison across various samples. This provides us with a test for our stellar parameter measurements while at the same time ensures that our newly derived abundances can be safely compared to our previous work. With this aim, we have re-derived the abundances for the Baade's Window sample of Red Giant Branch (RGB) stars from \citet{zoccali+08} and for the RC sample from \citet{hill+11} using the GALA-based procedure described above. Figure~\ref{parameters} shows the comparison between the stellar surface parameters and [Fe/H] for RC and RGB samples from \citet{hill+11} and those derived with our method. This comparison involves only the calculation of stellar parameters, since the EW measurement is done using DAOSPEC both in \citet{hill+11} and in this work.

As shown in Fig.~\ref{parameters}, there is a good agreement between both methods with no significant differences for either of the two samples analysed (RGB and RC stars).  In particular, for effective temperature we found a mean difference $\rm \Delta T_{eff} = 2 \pm 199$\,K and $-23 \pm 185$\,K for RGB and RC, respectively.  On the other hand, micro-turbulence velocities result in a mean difference of $\rm \Delta \xi=- 0.14\pm 0.19$ km/s for RGB stars and $\rm \Delta \xi= -0.12\pm 0.26$ km/s for RC stars. These stellar parameters translate into a mean metallicity difference of $\rm [Fe/H]=- 0.02\pm 0.19$ and $\rm [Fe/H]=- 0.01\pm 0.15$ dex. The derived differences are negligable and the scatter between both samples is in good agreement with the expected errors on the measurements of the surface stellar parameters as discussed in \citet{hill+11}. Furthermore, this confirms that our results and those from \citet{hill+11} are in good agreement, with no systematic differences between metalicities obtained for RGB stars from \citet{zoccali+08} and for RC stars as previously reported in \citet{hill+11}. Therefore, the measurements on the Baade’s Window samples (RC and RGB) can be compared in a consistent and homogenous way with the results from the GIBS survey discussed in the present study.

\begin{figure}[]
\centering
\includegraphics[width=9.0cm, angle=0,trim=0cm 0.0cm 0cm 0cm,clip=true]{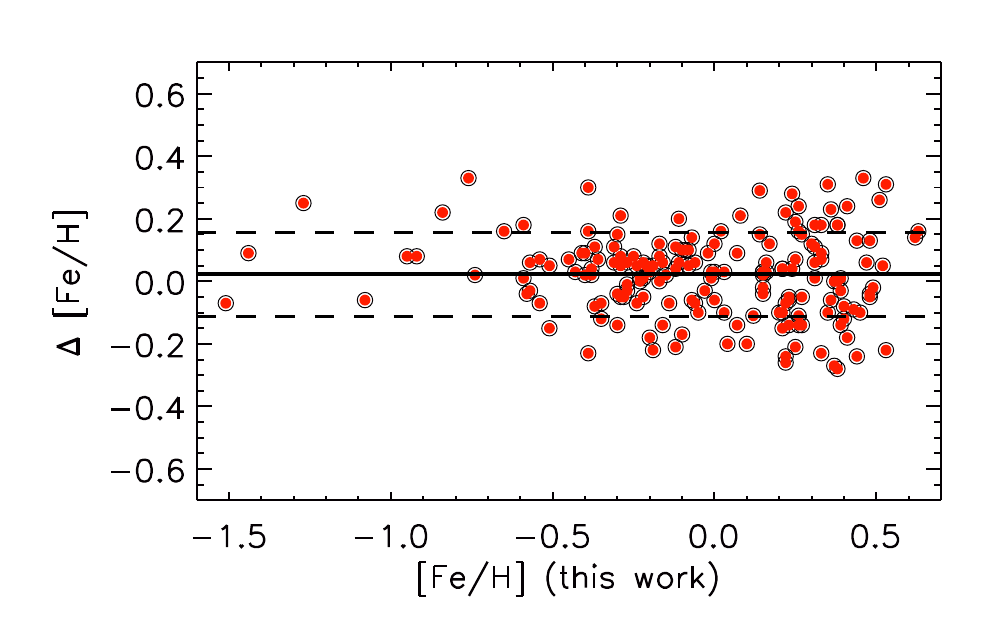}
\caption{Difference between the [Fe/H] abundances obtained with the adopted GALA procedure using the correct set of initial stellar surface parameters and using fixed values ($\rm T_{eff}$=4500 K, $\rm \xi$=1.5 km/s, $\rm \log{g}$=1.9, and [Fe/H]=0.00) for the RC sample from \citet{hill+11}}
\label{samepars}
\end{figure}

Finally, we investigated the dependence of the final [Fe/H] abundance on the initial photometric estimations of stellar parameters. Reddening corrections and distance spread translate into uncertainties in the initial photometric parameters. Furthermore, adopting an effective temperature calibration different to the one of \citet{ramirez+00} would cause a different starting point for the first run of GALA. In order to evaluate the impact of these parameters in the final abundances, we have tested a run of the RC sample from \citet{hill+11}, but starting from the same set of stellar parameters for all the stars: $\rm T_{eff}$=4500\,K, $\rm \xi$=1.5 km/s, $\rm \log{g}$=1.9 dex, and [Fe/H]=0.00. Figure~\ref{samepars} shows the comparison between the original [Fe/H] for this sample and the one resulting from the same stellar parameters. Because the mean difference in the [Fe/H] estimates is $\rm \Delta [Fe/H]$=0.02 dex, no significant systematic offsets are expected in the final set of [Fe/H] abundances due to the adopted set of initial stellar parameters, no matter how far they are from the correct values. On the other hand, the dispersion between the [Fe/H] is found to be $\rm \sigma_{[Fe/H]}$=0.13 dex, which is not negligible. Differences of up to 0.3 dex can be found among the most metal-rich stars as a result of using an incorrect set of initial parameters. It is worth mentioning that GALA has an optional initial stage, called the \textit{guess working block}, where the stellar parameter space is explored in a coarse grid in order to verify and refine poorly known initial parameters. However, in our case concerning Bulge giant stars, the reddening and distance spread uncertainties are always propagated into the calculation of the photometric surface gravity. The non-variation of this value against the procedure of GALA makes the \textit{guess working block} redundant and an error similar to the $\rm \Delta [Fe/H]$=0.13 dex observed in this test is expected within the uncertainty of this particular parameter. Most likely, the observed uncertainty associated to the initial stellar parameters is not caused by GALA itself, but for the inability to optimise the entire grid of parameters simultaneously in the bulge. Were a larger number of FeI and Fe II lines available in our spectra, the final scatter on stellar parameter would converge to less than 50 K in $\rm T_{eff}$ and 0.09 in $\rm \log{g}$ as described
in \citep{GALA+13} thus decreasing the relatively large scatter in [Fe/H] observed in Fig.~\ref{samepars}.

The uncertainties in the stellar parameters are derived internally by GALA. As pointed out in \citet{GALA+13}, the errors in the stellar parameters are dominated by the number of FeI lines used and their distribution in both $\chi_{\rm ex}$ range and transition values. For each stellar parameter, an optimisation parameter is calculated by applying a jackknife boot-strapping technique. This means that if the optimisation is computed using a set of N spectral lines, the parameters will then be re-calculated N times, but using N-1 lines each time. This calculation is then related to a parametrisation factor that measures the way each parameter affects the slopes and iron abundance differences within the GALA procedure. This parametrisation value is calculated by GALA by varying the best value found around the local minimum. This method and their respective equations are described in detail in \citet{GALA+13}. In our sample, the resulting errors are found to be $\rm \sigma_{T_{eff}}$=$245$ $\pm 99$ K, $\rm \sigma_{\xi}$=$0.29$ $\pm 0.11$ km/s. Each parameter is then varied by its corresponding uncertainty, while keeping the others unchanged, and the abundances are re-derived. The individual effects in abundance from the variations of each parameter are added in quadrature to obtain the final error in  metallicity. In our sample, this computation results in a mean metallicity error of $\rm \sigma_{[Fe/H]}$=$0.20 \pm 0.07$ dex. However, note that the errors in [Fe/H] can reach up to 0.4 dex for the most metal-rich stars but can be lower than 0.1 dex for metal-poor stars and are mostly driven by $\rm T_{eff}$ uncertainties. These errors are consistent with those reported in \citet{hill+11} and \citet{zoccali+08} for similar datasets and are also in agreement with the differences seen in Fig.~\ref{parameters}. Furthermore, the larger errors in stellar parameters towards the high metallicity regime are expected. Extensive analysis of errors in derivation of stellar parameters was presented in Smiljanic et al. (2014) in the context of the Gaia-ESO survey and showed how the errors become larger at high metallicities due to the increase of line blends.

\section{Alpha element abundances}

\begin{figure}[]
\centering
\includegraphics[width=9.0cm, angle=0,trim=0cm 0.0cm 0cm 0cm,clip=true]{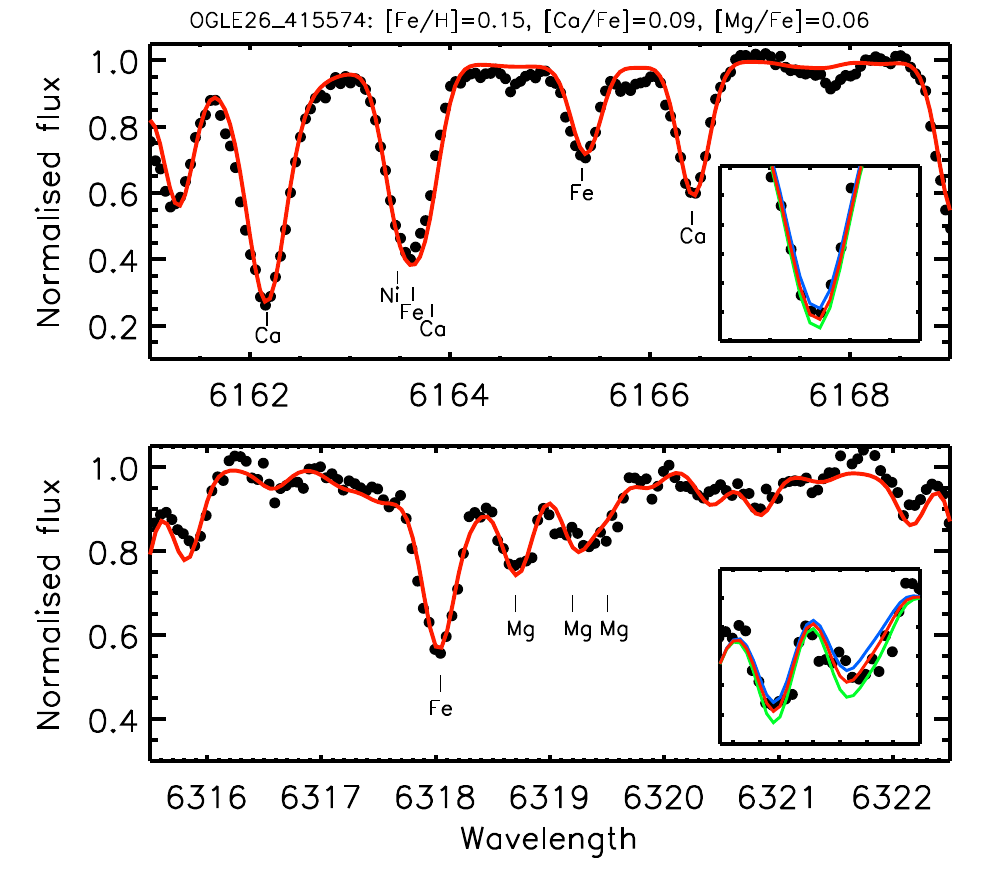}
\caption{Observed (black circles) and synthetic spectra for a star (OGLE26-415574) in the field HRm5m3 zoomed into the region used to derive Ca (upper panel) and Mg (lower panel) abundances. The synthetic spectra created with the best fitting Ca ([Ca/Fe]=0.06) and Mg ([Mg/Fe]=0.05) abundances are shown in solid red lines. The elements producing the stronger absorption lines are marked in each panel. For each spectral region we show a zoomed-in section around the Ca line at 6166.4 $\AA$ and the Mg triplet that includes the synthesis produced using the best-fitting abundance as red solid lines. In blue and green we show the synthetic spectra using a 0.1 dex abundance variation around the best-fitting value. The Y-axis of both zoomed-in regions have the same scale with normalized fluxes that range from 0.8--0.5 for Ca and 1.0--0.7 for Mg lines}.
\label{spec_region}
\end{figure}

In order to measure $\alpha$-element abundances for all the target stars we adopt the same procedure and line lists used in \citet{gonzalez+11}. As a first step, the corresponding ATLAS9 model atmosphere (Castelli \& Kurucz 2004) is generated using the stellar parameters found by GALA for each star. This model is then fed to MOOG \citep[v. Feb2013,][]{sneden+73}, in order to generate a synthetic spectrum. The synthetic spectrum is iteratively compared with the observed one after a local normalisation of the continuum, by varying only the elemental abundance of interest in each iteration until the best fitting abundance if found. 

\begin{table}
\begin{center}
\caption{Atomic line list for Mg and Ca using in this work. Also listed are the excitation potential ($\chi_{\rm ex}$) and oscillator strength ($\log gf$) for each analysed line.\label{lines}}
\begin{tabular}{c c c c c}
\\[3pt]
\hline
$\lambda$ (\AA) & Element & $\log gf$ & \large{$\chi_{\rm ex}$}\\
\hline
  6120.246 & FeI & -5.970 & 0.915\\
  6137.691 & FeI & -1.375 & 2.588\\
  6151.617 & FeI & -3.312 & 2.176\\
  6157.728 & FeI & -1.160 & 4.076\\
  6162.160 & CaI & -2.720 & 1.899\\ 
  6165.360 & FeI & -1.470 & 4.143\\
  6166.430 & CaI & 1.142  & 2.521\\
  6173.334 & FeI & -2.880 & 2.223\\
  6180.203 & FeI & -2.617 & 2.727\\
  6187.989 & FeI & -1.620 & 3.943\\
  6191.558 & FeI & -1.416 & 2.433\\
  6200.312 & FeI & -2.405 & 2.608\\
  6213.430 & FeI & -2.481 & 2.223\\
  6219.281 & FeI & -2.434 & 2.198\\
  6226.734 & FeI & -2.120 & 3.883\\
  6229.226 & FeI & -2.805 & 2.845\\
  6230.722 & FeI & -1.279 & 2.559\\
  6246.318 & FeI & -0.805 & 3.602\\
  6252.555 & FeI & -1.727 & 2.404\\
  6253.829 & FeI & -1.299 & 4.733\\
  6271.278 & FeI & -2.703 & 3.332\\
  6301.500 & FeI & -0.718 & 3.654\\
  6311.500 & FeI & -3.141 & 2.831\\
  6318.710 & MgI & -2.000 & 5.110 \\ 
  6319.230 & MgI & -2.240 & 5.110 \\ 
  6319.490 & MgI & -2.680 & 5.110 \\  
  6322.685 & FeI & -2.448 & 2.588\\
  6330.848 & FeI & -1.64 & 4.733\\
  6335.330 & FeI & -2.177 & 2.198\\
  6336.823 & FeI & -0.856 & 3.686\\
  6355.028 & FeI & -2.32 & 2.845\\
  6380.743 & FeI & -1.475 & 4.186\\
\hline
\end{tabular}
\end{center}
\end{table}

Based on similar dataset -- same setup, resolution and similar signal-to-noise -- \citet{gonzalez+11} showed that among all the $\alpha$-elements measurable in the observed spectral range Mg has the smallest scatter. Therefore, we consider Mg to be the key element allowing a precise comparison between the abundance trend observed in different bulge and disc star samples. However, in order to obtain precise abundances of Mg we need to first calculate Ca abundances. Indeed, as discussed in detail in \citet{lecureur+07}, \citet{hill+11}, and \citet{gonzalez+11} the continuum around the Mg triplet at $6319 \AA$ is severely affected by $6318.1 \AA$ Ca I autoionisation line. The large broadening of this Ca line causes a drop in the Mg triplet region that can result in an incorrect abundance measurement for Mg. For this reason, we first measured Ca abundances using two lines at  6162.1 and 6166.43$\AA$ (see Fig.~\ref{spec_region}, lower panel and Table~\ref{lines}). Once the Ca abundances have been measured we provide this value as an additional input for MOOG to be included in the synthesis used for the calculation of Mg abundances. The local normalisation of the continuum is then manually improved as an additional way to account for any failure to reproduce the underlying Ca autoionisation line. An example of this fit is shown in Fig.~\ref{spec_region}.  This procedure was applied to all the analysed stars, obtaining Ca and Mg abundances for a total of 400 stars spread in our four fields. Fig.~\ref{spec_region} also shows an example of the sensitivity of the lines to the variations in [Mg/H] and [Ca/H].  For a variation of $\pm 0.1$ dex, the effect in the fit region of Mg is clearly seen, while the Ca lines appear to be less sensitive to such variations. Both of the Ca lines that are used in this work show a similar behaviour. The difference in [Ca/Fe] abundances measured from each of the lines has a $\sigma$ of 0.05 dex, which is negligible compared to the errors involved in our analysis due to the stellar parameters uncertainties.

\begin{figure}[]
\centering
\includegraphics[width=9.3cm, angle=0,trim=0cm 0.0cm 0cm 0cm,clip=true]{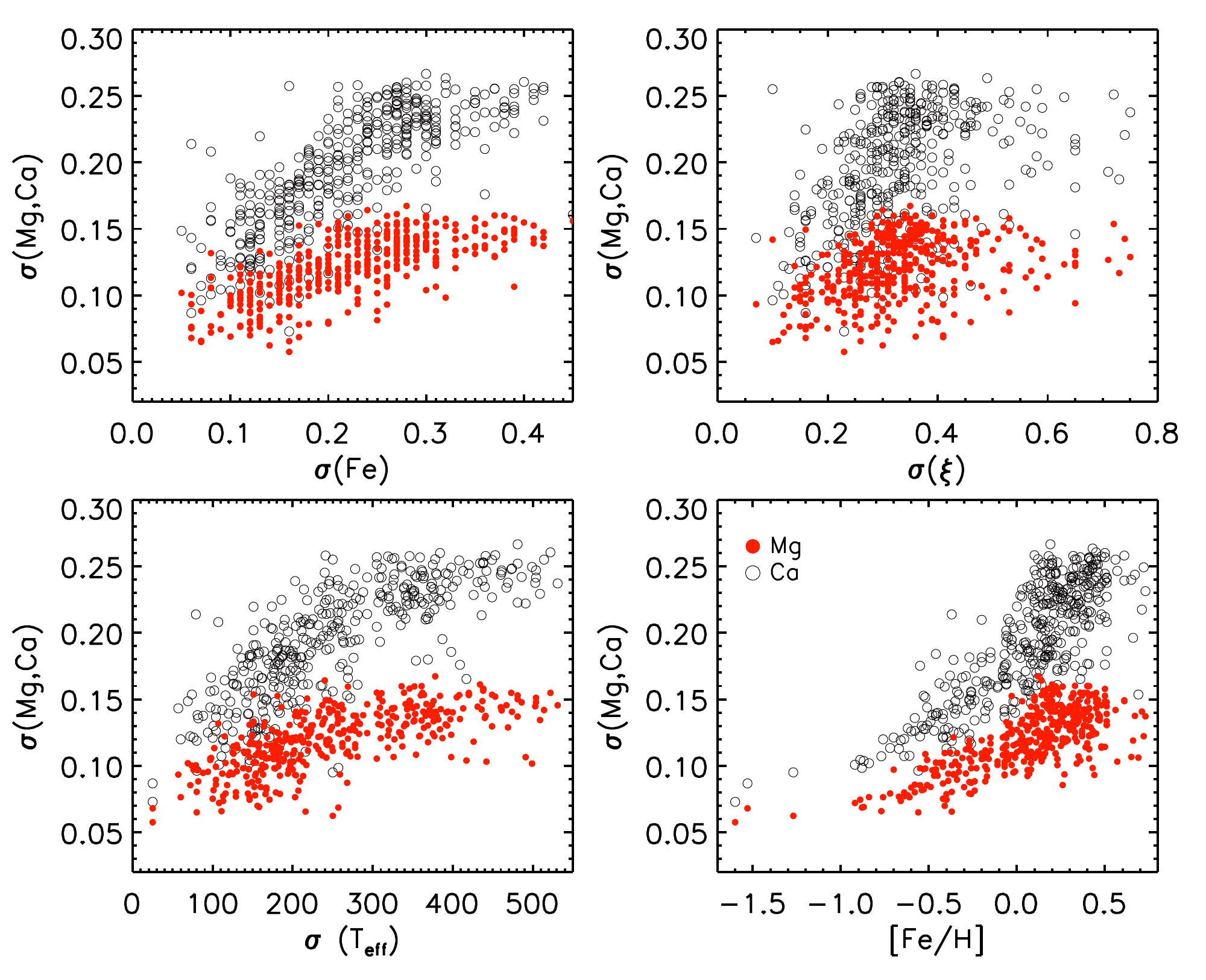}
\caption{Abundance measurement errors as a function of errors in the stellar parameters: [Fe/H] (upper left panel), microturbulence velocity (upper right panel) and effective temperature (lower left panel). Errors as a function of the corresponding [Fe/H] value are also displayed (lower right panel). In all panels errors for [Ca/H] abundances are shown as red empty circles and errors for [Mg/H] as red filled circles.}
\label{errors}
\end{figure}

The uncertainties on the measured [Mg/H] and [Ca/H] abundances were calculated in the same way as in \citet{gonzalez+11} by varying the stellar parameters by their corresponding uncertainties and re-calculating the best fitting abundances. These errors are then added in quadrature together with the error from our spectral synthesis fitting procedure of 0.1 dex derived in \citet{gonzalez+11}. The errors are individually reported in Table~\ref{allstars} and are found to be of the order of 0.14 dex for [Mg/H] and 0.21 dex for [Ca/H]. Figure~\ref{errors} shows the error estimation for Ca and Mg abundances as a function of the estimated uncertainties for the stellar parameters of each star. It can be clearly seen that Ca abundances are much more sensitive to any variations on stellar parameters than Mg abundances. Furthermore, the larger sensitivity of Ca abundances to stellar parameters is seen at all metallicities, but is even larger in the metal-rich regime. 

\section{Results}

\begin{figure}[]
\centering
\includegraphics[width=9cm, angle=0,trim=0cm 2.5cm 0cm 0.8cm,clip=true]{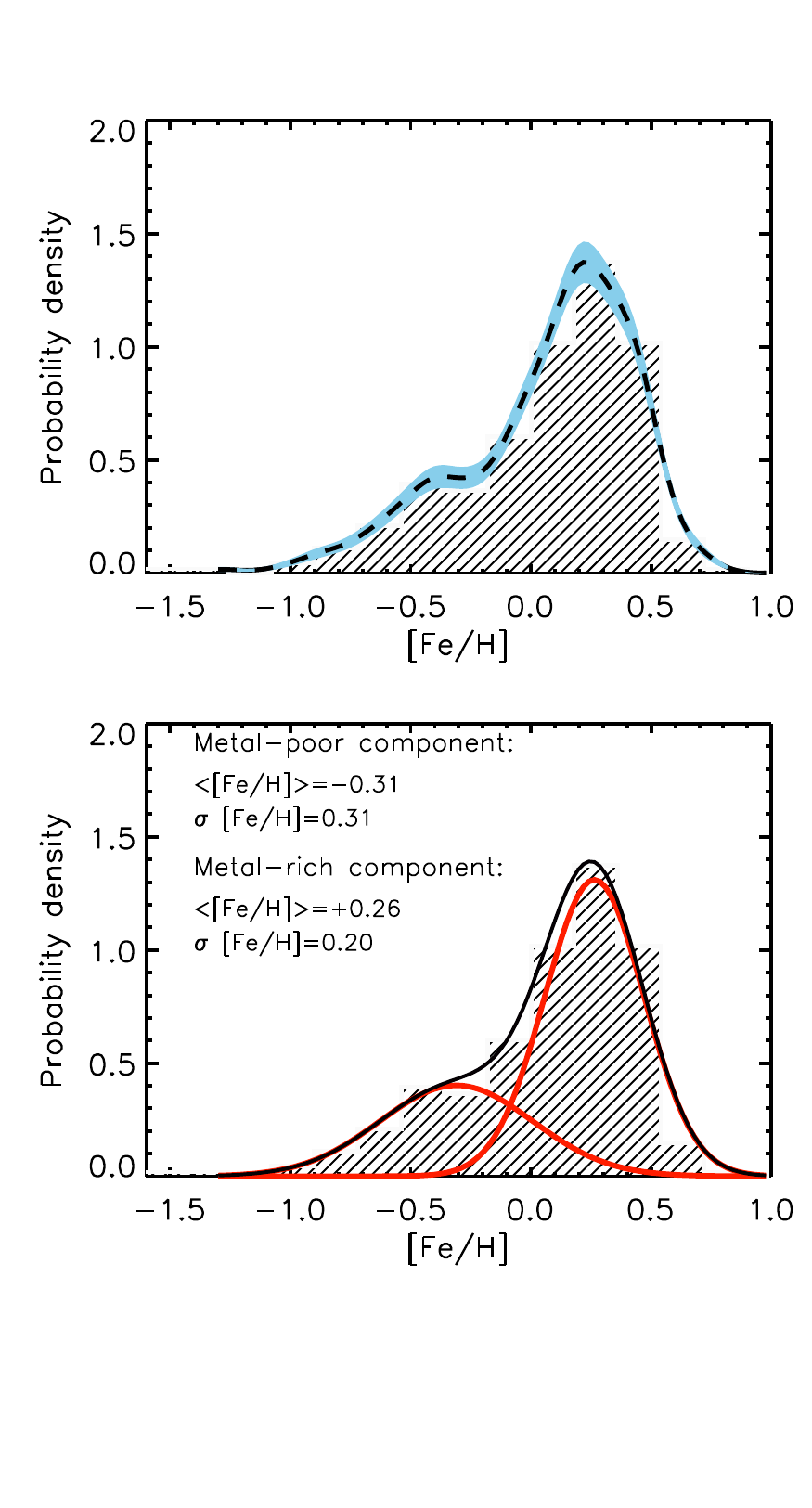}
\caption{Overall metallicity distribution obtained from the combination of the 4 GIBS high-resolution fields and the RC sample of Hill et al. (2011). The probability density distribution is overplotted in the upper panel as a dashed line. The corresponding variance on the calculated probability densities is shown in blue. The lower panel shows the best two Gaussian fit to the upper panel distribution. The resulting distribution from the two Gaussians is shown as a solid line.}
\label{all_fe}
\end{figure}

Stellar parameters, [Fe/H], [Ca/Fe] and [Mg/Fe] values measured for a total of 400 stars in the GIBS sample are listed in Table~\ref{allstars} and discussed in detail in this section.

\begin{table*}
\begin{center}
\caption{Stellar parameters, [Fe/H], and [Mg/Fe], [Ca/Fe] abundance ratios for stars in the four high-resolution GIBS fields.  The full table is available in electronic form. Individual coordinates and the associated photometry is available in \citet{zoccali+14}.\label{allstars}}
\begin{tabular}{c c c c c c c c c c c c c}
\hline\hline
  \multicolumn{1}{c}{Field} &
  \multicolumn{1}{c}{STAR} &
  \multicolumn{1}{c}{$\rm T_{eff}$} &
  \multicolumn{1}{c}{$\rm \sigma T_{eff}$} &
  \multicolumn{1}{c}{$\rm \xi$} &
  \multicolumn{1}{c}{$\rm \sigma \xi$} &
  \multicolumn{1}{c}{$\rm \log{g}$} &
  \multicolumn{1}{c}{[Fe/H]} &
  \multicolumn{1}{c}{$\rm \sigma [Fe/H]$} &
  \multicolumn{1}{c}{[Ca/Fe]} &
  \multicolumn{1}{c}{$\rm \sigma [Ca/Fe]$} &
  \multicolumn{1}{c}{[Mg/Fe]} &
  \multicolumn{1}{c}{$\rm \sigma [Mg/Fe]$} \\
\hline
HRm5m3 &  OGLE26\_649743 & 4328 & 240 & 1.6 & 0.30 & 2.29 & 0.14 & 0.24 & 0.02 & 0.21 & 0.14 & 0.09\\
HRm5m3 &  OGLE26\_661709 & 4537 & 269 & 1.4 & 0.25 & 2.22 & 0.28 & 0.33 & 0.01 & 0.19 & -0.04 & 0.09\\
HRm5m3 &  OGLE26\_649746 & 4501 & 257 & 1.0 & 0.24 & 2.36 & 0.35 & 0.21 & 0.15 & 0.20 & 0.01 & 0.16\\
HRm5m3 &  OGLE26\_661748 & 4468 & 151 & 1.3 & 0.28 & 2.52 & 0.24 & 0.19 & 0.00 & 0.22 & 0.01 & 0.21\\
         ...... &                 ......         &  ......   &  ......& ...... &  ......  &   ...... &   ...... &  ...... &  ......  &  ...... &  ......  &  ...... \\
\hline
\end{tabular}
\end{center}
\end{table*}

\subsection{The Bulge metallicity distributions at constant latitude}

\begin{figure*}[]
\centering
\includegraphics[width=18.2cm, angle=0]{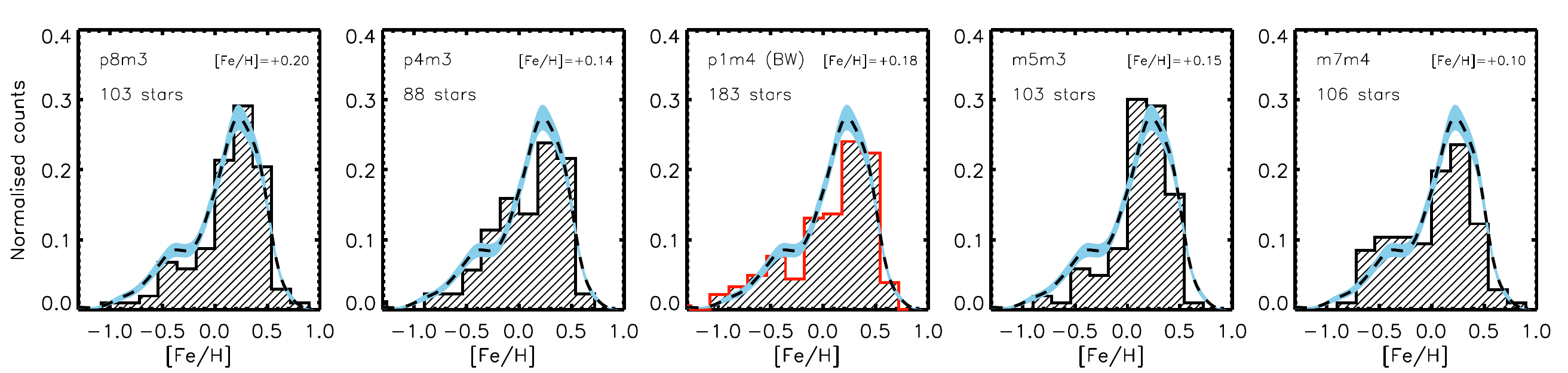}
\caption{Metallicity distributions for the four GIBS high-resolution fields. Additionally, the metallicity distribution for Baade's window from Hill et al. (2011) is shown in red in the middle panel. For each of the metallicity distributions we overplot the probability density distribution for the entire sample of metallicities shown in Fig.~\ref{all_fe}.} 
\label{mdfs}
\end{figure*}

The derived metallicity distribution of the five fields located at latitude b$\sim-3.5^{\circ}$ (i.e. the 4 GIBS fields and the Baade's window RC sample from \citet{hill+11}) are very similar to each other (see Fig.~\ref{mdfs}). Therefore, in order to increase the number statistics, we initially construct the global metallicity distribution from all the stars in the 5 fields together. Fig.~\ref{all_fe} shows the resulting global metallicity distribution. In order to avoid binning effects when looking at the shape of the distribution, we have estimated the probability density underlying in the global sample by using the kernel density estimator method (Silverman et al. 1998). We have adopted a Gaussian kernel and the optimal smoothing parameter as defined in Silverman et al. (1998). 

The good number statistics of the total sample allows us to investigate in detail the metallicity distribution. In particular, we are interested in evaluating the decomposition of the distribution in two populations, as suggested in previous studies \citep{hill+11, rojas-arriagada+14}. Note that \citet{ness-abu+13} suggested that the Bulge metallicity distribution can be decomposed into five Gaussian components, each of them corresponding to different populations from the bulge and foreground disc. Here, we limit this decomposition to only two Gaussian components. The main reason for setting this limit is because studies, based on both bulge kinematics and morphology, have found evidence for only two main components: a metal-poor spheroid-like component and a metal-rich bar-like component. Also, in this study we are restricted to inner Bulge fields in which we have a narrow target selection box in the colour-magnitude diagram. For this reason, the contamination by the thin/thick disc and halo is negligible here \citep[components D, F, and E in][]{ness-abu+13}. It is worth stressing that here the use of two Gaussian distributions is not necessarily meant to describe the actual shape of the metallicity distribution of each of the two components. It represents the attempt to perform the best possible parametrisation of the observed global metallicity distribution, so we can therefore investigate how its shape changes across the fields.

We perform a least-squares Gaussian fit to the global metallicity distribution using the IDL routine XGAUSSFIT \citep{Lindler+2001}. The code automatically fits the main Gaussian component, and additional components can be then included by specifying an initial guess for the mean, peak, and sigma value. The code then refines the fit for both Gaussian functions simultaneously. The best-fit Gaussian distributions and the global metallicity distribution are shown in Fig.~\ref{all_fe}. The resulting parameters for the best-fit metal-poor Gaussian component are $\rm <[Fe/H]>$=$-0.31$ and $\rm \sigma [Fe/H]$=$+0.31$. On the other hand, the metal-rich Gaussian component has a mean $\rm <[Fe/H]>$=$+0.26$ and $\rm \sigma [Fe/H]$=$+0.20$. The two Gaussian fits to our global metallicity distribution are in good agreement with those of \citet{hill+11} who finds a metal-poor component with mean of $\rm [Fe/H]$=$-0.27$ and a metal-rich one of $\rm [Fe/H]$=$+0.29$. The sigma of the metal-poor component is in good agreement with that of \citet{hill+11}, finding a broad metal-poor component with $\rm \sigma [Fe/H]$=$0.31$. On the other hand, \citet{hill+11} find a metal-rich component slightly narrower than ours with $\rm \sigma [Fe/H]$=$0.12$. Nevertheless, the results are fully compatible. The observed difference in the width of the metal-rich components originates from the increase of the [Fe/H] errors towards the metal-rich regime and the respective error deconvolution that was applied in the \citet{hill+11} analysis. 

The individual metallicity distributions for all the fields presented in this work are shown in Fig.~\ref{mdfs}. From the measured mean metallicity in each field we see no evidence of a gradient such as the one seen along the minor axis \citep{zoccali+08, ness-abu+13}. The observed mean [Fe/H] values of the fields studied in this work range between [Fe/H]=0.10--0.20, consistent with what is shown in the mean photometric metallicity map from \citet{gonzalez+13}. We do not attempt a multi-Gaussian fit to the individual field metallicity distributions because we do not have enough statistics in each field, therefore preventing us from deriving a statistically robust conclusion. However, we use the probability density distribution that was obtained from the global metallicity distribution as a reference in order to evaluate field-to-field variations. From Fig.~\ref{mdfs} we notice that the metallicity distribution of individual fields span the same range: $\rm -1 \leq [Fe/H] \leq +1 $. The general shape of the distributions is well represented by the probability density distribution constructed over the global metallicity distribution. A more noticeable difference can be observed in field m5m3 where the metal-rich side of the distribution shows a strong peak near Solar metallicity that is not seen in the other fields.

In order to evaluate whether the metallicity distributions for each field originate from different populations or not, we performed a two sample Kolmogorov-Smirnov (K-S) test and compared the metallicity distribution found in Baade's window with the individual GIBS field distributions. The K-S test returns p-values (i.e. the probability that two functions belong to the same distribution): 0.22 for p8m3, 0.48 for p4m3, 0.03 for m5m3, and 0.16 for m7m3. With the exception of field m5m3, the K-S test confirms that our samples are most likely drawn from the same parent population. For field m5m3, the result is not conclusive. From the visual inspection of the metallicity distribution we notice that the main difference with respect to the other fields comes from a concentration of stars at [Fe/H]$\sim$+0.1. 

We conclude in favour of a null mean metallicity gradient along the Bulge major axis, as previously suggested in \citet{johnson+13} at higher latitudes (but see \citet{rangwala+09} and \citet{babusiaux+14}). Furthermore, we conclude that the metallicity distributions presented in this work for fields along the longitude strip at b=$-4^{\circ}$ originate from the same parent population. Based on this, we suggest that the relative contribution of the potentially two different Bulge components (spheroid-like and bar) does not change along the major axis, providing that this is indeed the origin of the metallicity gradient found along the minor axis.

\subsection{The Bulge alpha-element abundances at constant latitude}

\begin{figure}[]
\centering
\includegraphics[width=9.0cm, angle=0,trim=0cm 0.0cm 0cm 0cm,clip=true]{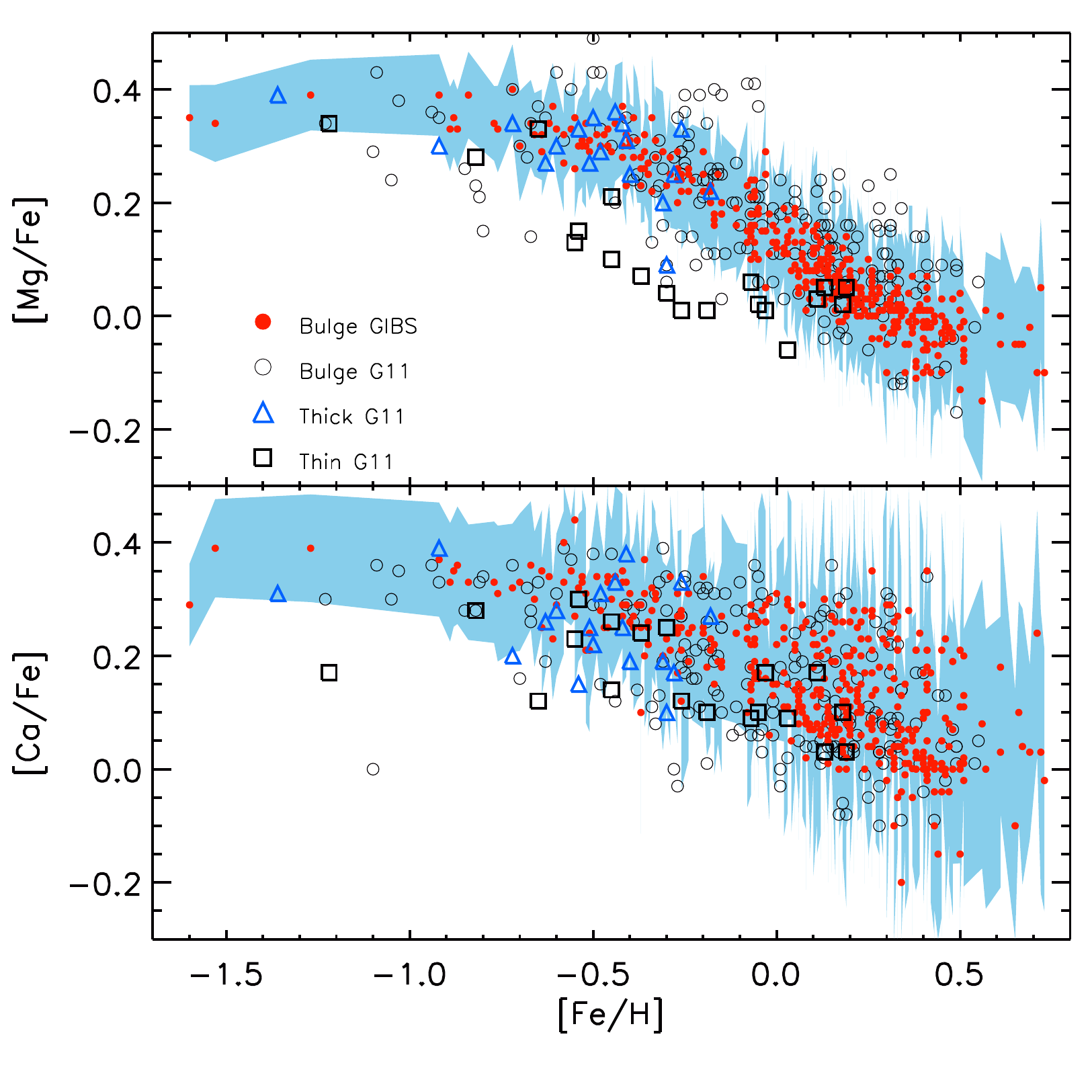}
\caption{Distribution of [Mg/Fe] (upper panel) and [Ca/Fe] (lower panel) as a function of [Fe/H] used as a diagnostic of the formation time-scale of the Bulge. Shown as red circles are the [Mg/Fe] and [Ca/Fe] values obtained from the HR13 spectra of 400 stars in the four GIBS fields presented in this work compared to those for RGB stars from the Baade's window sample of Gonzalez et al. (2011) which are shown as black empty circles. Measurements for the local thin (black squares) and thick (blue triangles) disc stars from the sample from \citet{gonzalez+11} are also shown in both panels. The uncertainties for [Ca/Fe] and [Mg/Fe] abundances are shown as a light-blue contour in both panels according to the individual uncertainties reported in Table~\ref{allstars}.}
\label{Camg}
\end{figure}

We now present the results for [Ca/Fe] and [Mg/Fe] abundances for all our target stars, obtained by comparison to synthetic spectra created using the corresponding stellar surface parameters and [Fe/H] discussed in previous sections. A total sample of 400 stars have been analysed and their [Ca/Fe] and [Mg/Fe] over [Fe/H] trends are shown in Fig.~\ref{Camg}. Also in Fig.~\ref{Camg} we show a comparison of our results with respect to those derived from a similar sample in Baade's window measured in \citet{gonzalez+11}. 

\begin{figure}[]
\centering
\includegraphics[width=9.0cm, angle=0,trim=0cm 0.0cm 0cm 0cm,clip=true]{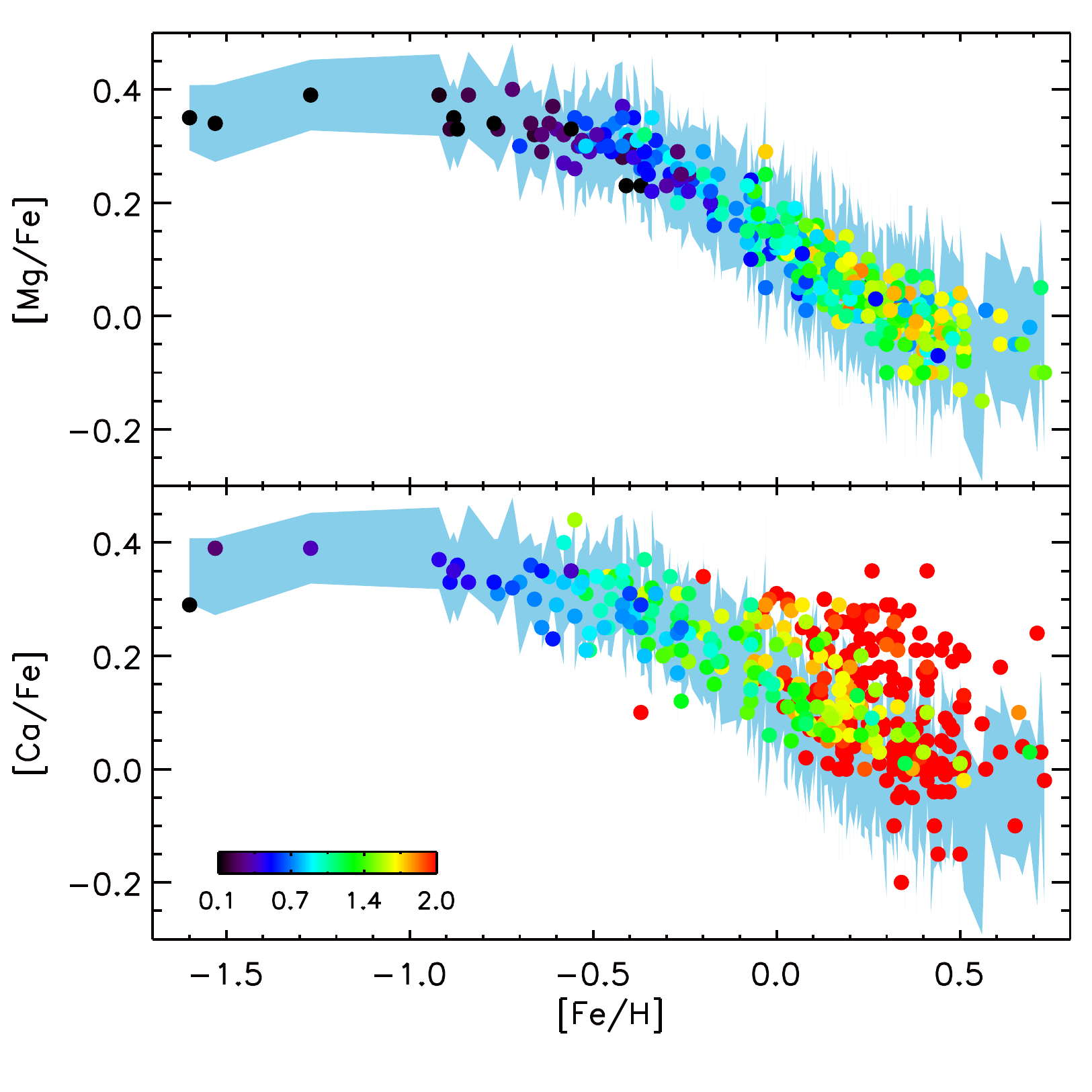}
\caption{Distribution of [Mg/Fe] (upper panel) and [Ca/Fe] (lower panel) as a function of [Fe/H] as in Fig.~\ref{Camg_fields}. Here we colour-coded the abundances of [Ca/Fe] and [Mg/Fe] according to the estimated uncertainty value. The colour-coding was fixed in both panels to the range of [Mg/Fe] uncertainties between 0.07 -- 0.20 dex from black to red. The uncertainty contour of [Mg/Fe] is also shown in the background of both panels as a reference.}
\label{errors_color}
\end{figure}

While the scatter of the measurements in [Ca/Fe] are comparable in both samples, the [Mg/Fe] trend shows a remarkable reduction of the scatter with respect to the previous measurement from \citet{gonzalez+11}. This is most likely a consequence of our improved calculation of stellar parameters as discussed in Section 3. The larger dispersion in Ca abundances with respect to Mg can be clearly seen in Fig.~\ref{Camg}. This spread in [Ca/Fe] is most likely caused by the larger dependence of Ca abundance on stellar parameter uncertainties. In Fig.~\ref{errors_color} we show that when only the stars with small uncertainties are considered, the sequence of [Ca/Fe] becomes tighter, with much less scatter at all metallicities. It can be seen that stars with high [Ca/Fe] found at high metallicities [Fe/H] $>$ 0 are restricted to those measurements with large uncertainties. In our measurements any error in Ca abundances should in principle translate in an error in Mg abundances due to the underlying auto-ionisation Ca line that strongly affects the continuum around the Mg triplet \citep{hill+11}. Therefore, the adoption of an incorrect [Ca/Fe] abundance would result in a larger scatter for [Mg/Fe] abundances due to a poorer quality fit in the respective region. We do not see this large spread in Mg abundances, most likely because this is partially solved by the local normalisation of the continuum. As a matter of fact, the resulting [Mg/Fe] trend for the Bulge can be traced with a great level of detail. 

We compare the trends of [Ca/Fe] and [Mg/Fe] as a function of [Fe/H] for our entire sample with those from giant stars of the local thin and thick disc published in \citet{gonzalez+11}. As seen in both Ca and Mg abundances in Fig.~\ref{Camg}, the $\alpha$-element enhancement of the Bulge with respect to the thin disc is found at all metallicities. The thick disc on the other hand appears as enhanced in $\alpha$-element abundances as the metal-poor Bulge giants, though covering a much smaller range in [Fe/H]. These patterns in the $\alpha$-element abundances of Bulge stars with respect to the disc abundances has been already observed in several studies, based on different samples and measurement techniques \citep{melendez+08, alves-brito+10, gonzalez+11, bensby+10, bensby+11, ryde+10, hill+11, johnson+11, johnson+13, johnson+14}. In addition, we have calculated the metallicity at which the [Mg/Fe] abundances starts to decrease (the so-called \textit{knee}) similarly as in  \citet{gonzalez+11}, by producing a linear fit to the metal-poor ([Fe/H]=$-1.6$ to $-0.6$) and metal-rich ([Fe/H]=$-0.3$ to $+0.3$) stars. The intersection between both linear fits corresponds to the $\alpha$-element \textit{knee} located at [Fe/H]=-0.44, in good agreement with the values found in \citet{gonzalez+11} in a field located at the same latitude  b=$-4^{\circ}$.

\citet{bensby+13} suggested that the position of the \textit{knee} in [$\alpha$/Fe] plot is found at 0.1--0.2 dex higher metallicities in the Bulge than in the thick disc. A similar result was proposed by \citet{johnson+14} who added the analysis of Fe-peak elements finding in particular that Co, Ni, and Cu appear enhanced compared to the disc. It is important to recall that the results presented in \citet{johnson+14} have been obtained by comparing Bulge giants to dwarf stars from the disc. This technique might suffer from systematic offsets \citep{melendez+08} and it thus needs to be further investigated. The results from \citet{bensby+13} do not suffer from these systematics, as the comparison involves only dwarf and sub-giant stars for both disc and bulge; however, because the microlensed dwarfs sample is still relatively small, for the time being the derived result is far from being statistically robust.  Unfortunately, in our work we do not have measured abundances for a sufficiently large number of thick disc giant stars to calculate the position of the corresponding [$\alpha$/Fe] \textit{knee} with enough accuracy to compare it with the value ([Fe/H]=-0.44\,dex) measured in the GIBS fields. However we note that for the Bulge we find a [$\alpha$/Fe] \textit{knee} $\sim$0.1 dex more metal-poor than the value presented in \citet{bensby+13}. This means that, assuming that our study is in the same scale as \citet{bensby+13}, our abundances would not favour a more metal-rich \textit{knee} for the Bulge than the local thick disc.

\begin{figure}[]
\centering
\includegraphics[width=9.0cm, angle=0,trim=0cm 0.0cm 0cm 0cm,clip=true]{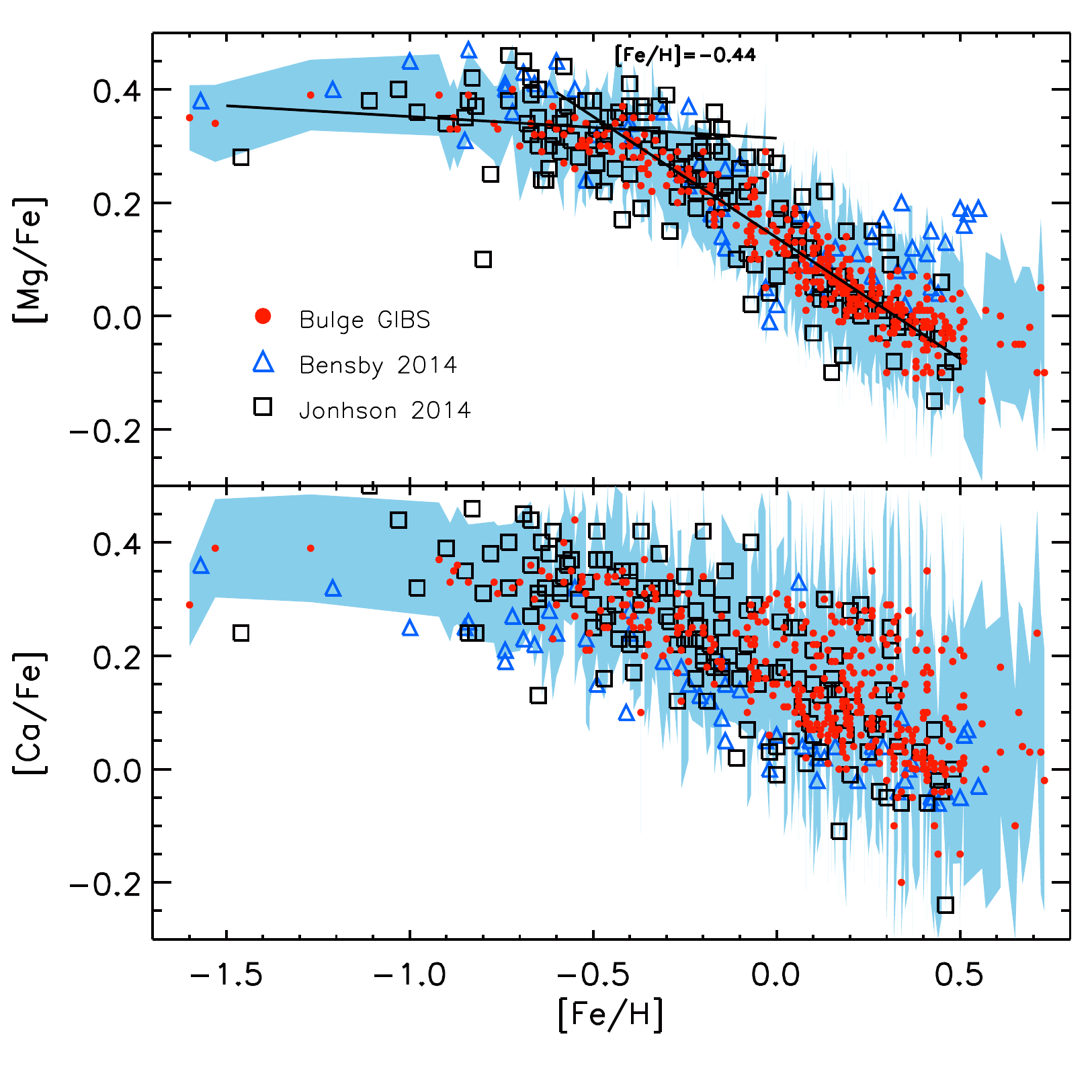}
\caption{Distribution of [Mg/Fe] (upper panel) and [Ca/Fe] (lower panel) as a function of [Fe/H] for the GIBS high-resolution fields studied in this work and for the measurements from \citet{johnson+14} for giant stars  (black squares) and microlensed dwarfs from \citet{bensby+13} (blue triangles). In the upper panel, the solid lines show the best-fit relation between [Mg/Fe] and [Fe/H] for the metal-poor and metal-rich regime. The intersection point is shown and represents the position of the \textit{knee}. The uncertainties for [Ca/Fe] and [Mg/Fe] abundances of this work are shown as a light-blue contour in both panels.}
\label{Camg_compare}
\end{figure}

In Fig.~\ref{Camg_compare} we show a comparison between the [Mg/Fe] and [Ca/Fe] trends between GIBS giants, the giants analysed in \citet{johnson+14}, and the microlensed dwarfs from \citet{bensby+13}. The trends are very similar for [Mg/Fe] abundances while for [Ca/Fe] there is a clear systematic offset of the \citet{bensby+13} abundances with respect to both GIBS and \citet{johnson+14}. 
The fact that the offset is not seen in [Mg/Fe] indicates that most likely it is not due to a different method of spectral analysis,
but it emphasizes the risk of comparing abundances for stars in very different temperature and/or gravity regimes (giants
versus dwarfs, in this case) because the abundances of some elements may be affected by different systematic errors. 

\begin{figure}[]
\centering
\includegraphics[width=9.0cm, angle=0,trim=0cm 0.0cm 0cm 0cm,clip=true]{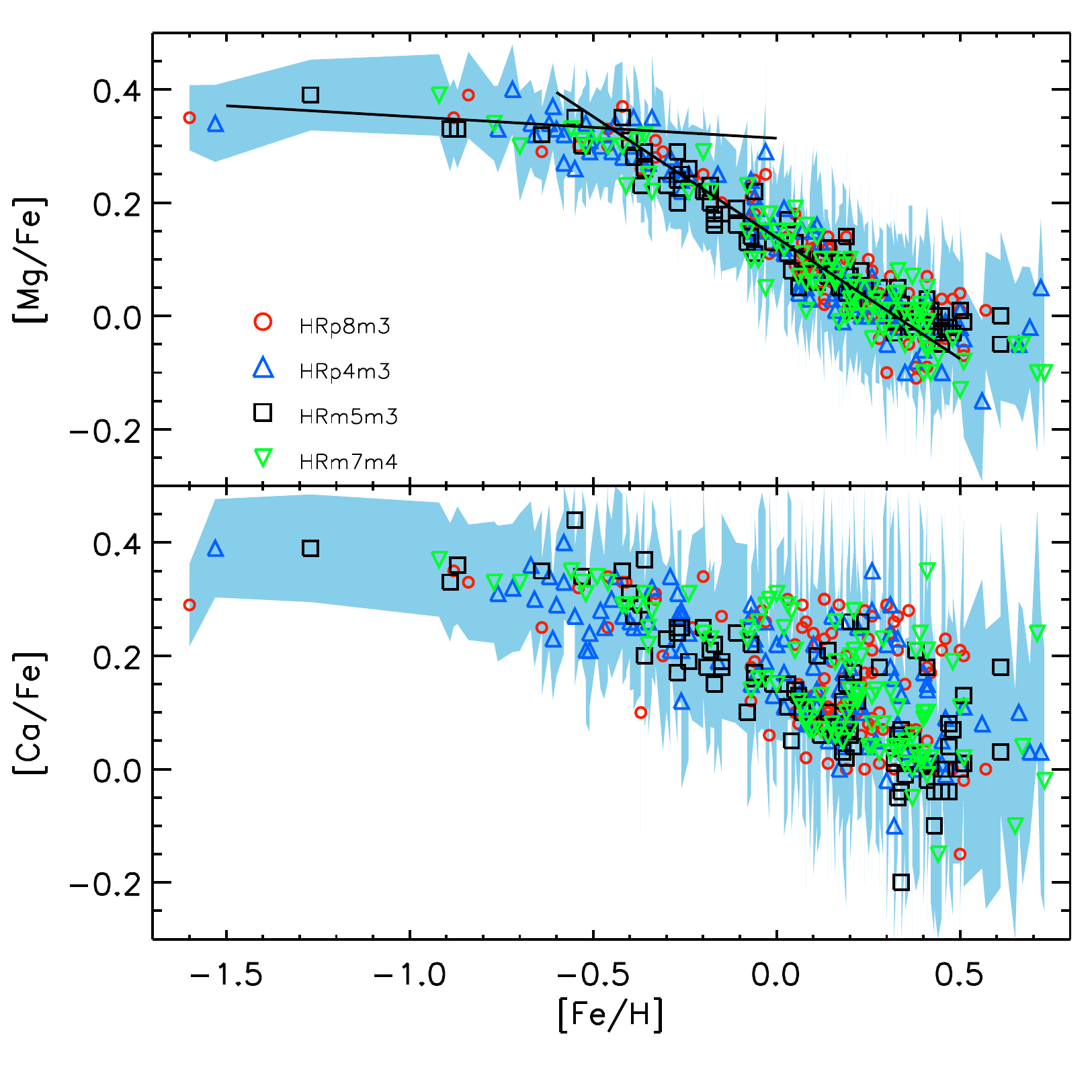}
\caption{Distribution of [Mg/Fe] (upper panel) and [Ca/Fe] (lower panel) as a function of [Fe/H] using different symbols for the four GIBS high-resolution fields studied in this work. The solid lines show the best-fit relation between [Mg/Fe] and [Fe/H] for the metal-poor and metal-rich regime obtained using the entire sample. The uncertainties for [Ca/Fe] and [Mg/Fe] abundances are shown as a light-blue contour in both panels.}
\label{Camg_fields}
\end{figure}

Finally, in Figure~\ref{Camg_fields} we show the individual trends of [Mg/Fe] and [Ca/Fe] abundances as a function of [Fe/H] for the four fields presented in this work, compared to the [Mg/Fe] abundances for the local thick and thin disc. No evident field-to-field variations can be seen and the same similarity between the metal-poor Bulge (i.e. [Fe/H]$\leq -0.5$) and the local thick disc is found in all fields. This result for a strip at constant Galactic latitude is in agreement with the conclusions of \citet{gonzalez+11} for the minor axis.

\section{Conclusions}

In this work we presented the surface stellar parameters, and the [Fe/H], [Ca/Fe], and [Mg/Fe] element ratios for 400 RC stars in four fields of the GIBS survey, corresponding to the high-resolution sample at $\rm b=-4^{\circ}$ observed with GIRAFFE HR13 setup. Spectroscopic stellar surface parameters were obtained with an automatic method based on the usage of the GALA code. These parameters were used to obtain [Ca/Fe] and [Mg/Fe] abundances for each star based on fitting of synthetic spectra. The global and individual field metallicity distributions were constructed, in order to investigate the presence of gradients and field-to-field changes in the shape of the distributions. We also investigated the [Ca/Fe] and [Mg/Fe] abundances as a function of [Fe/H] as a constraint for the time-scale of formation for the Bulge.

The results can be summarised as follows:

\begin{itemize}

\item A null metallicity gradient was found along Galactic longitude for a constant Galactic latitude. The metallicity distributions of all the fields presented here are consistent with being originated from the same parent population.

 \item A two Gaussian component provides a good fit to the global metallicity distribution at b=$-4^{\circ}$ with a narrow metal-rich ([Fe/H]$\sim$+0.26) population of stars and a broad metal-poor ([Fe/H]$\sim$-0.31) component. This is in agreement with previous studies suggesting a bimodal population for the Bulge. This exercise has been carried on because several independent observations, including stellar kinematics and distance distribution, suggest the presence of two stellar populations in the Galactic bulge. Whether the two components have indeed a different formation history, needs to be seen by model comparisons to observations as the ones presented here.

\item A very tight relation was found between [Mg/Fe] and [Fe/H], showing enhancement with respect to the solar ratio in the metal-poor regime, similar to the one seen in the local thick disc, and a decrease in [Mg/Fe] starting at [Fe/H]$\sim$-0.44. All studied fields share these properties, with no appreciable differences among each other. The observed position of the so-called \textit{knee} is in agreement with our previous results but it occurs at $\sim$0.1\,dex lower metallicity than what is observed in the bulge microlensed dwarf and sub-giants.

\item The [Mg/Fe] and [Ca/Fe] results presented here are fully compatible with previous results based on giant stars; however, we see a systematic offset in [Ca/Fe] with respect to the abundances from the microlensed dwarf stars in \citet{bensby+13}. This difference is not seen in [Mg/Fe]. This offset is most likely due to different systematics affecting stars in different temperature and gravity regimes.

\end{itemize}

\begin{acknowledgements}
We are grateful for the useful comments received from an anonymous referee. We warmly thank the ESO Paranal Observatory staff for performing the observations for this programme. MZ and DM acknowledge funding from the BASAL CATA through grant PFB-06, and the Chilean Ministry of Economy through ICM grant to the Millennium Instritute of Astrophysics. MZ acknowledge support by Proyecto Fondecyt Regular 1150345. DM acknowledge support by FONDECYT No. 1130196.
\end{acknowledgements}

\bibliographystyle{aa}
\bibliography{mybiblio_rev_full}

\end{document}